\documentclass[twocolumn,epsfig,floats,aps]{revtex4}
\usepackage{amsmath,bm}
\usepackage{epsfig}
\usepackage{graphicx}

\begin{document}
\title{PACIAE Model Predictions for $Pb+Pb$ Collisions at LHC Compared to the 
       $Au+Au$ Collisions at RHIC}
\author{Ben-Hao Sa$^{1,2,3}$ \footnote{sabh@ciae.ac.cn}, Dai-Mei Zhou$^2$, 
        Bao-Guo Dong$^1$, Yu-Liang Yan$^1$, Hai-Liang Ma$^1$, and Xiao-Mei Li$^1$}
\affiliation{
$^1$  China Institute of Atomic Energy, P. O. Box 275 (18),
      Beijing, 102413 China \\
$^2$  Institute of Particle Physics, Huazhong Normal University,
      Wuhan, 430079 China \\
$^3$  CCAST (World Lab.), P. O. Box 8730 Beijing, 100080 China}
\begin{abstract}
The parton and hadron cascade model PACIAE is used to simulate the 0-6, 15-25, 
and 35-45\% most central $Au+Au$ collisions at $\sqrt{s_{NN}}$=19.6, 62.4, 
130, and 200 GeV and the 0-10\% most central $Pb+Pb$ collisions at 
$\sqrt{s_{NN}}$=5500 GeV. The calculated charged multiplicity and the 
charged particle transverse momentum distribution, pseudorapidity distribution, 
and extended longitudinal scaling for $Au+Au$ collisions well agree with 
the corresponding PHOBOS data. Thus the above observables calculated for 
$Pb+Pb$ collisions would be a reliable prediction.   
\end{abstract}
\maketitle

In Ref. \cite{arm1} the predictions of variety models for $Pb+Pb$ collisions 
at LHC energy ($\sqrt{s_{NN}}$=5500 GeV) have been compiled including the 
thermal (statistical) models \cite{my}. The basic assumption in thermal model 
is that the final state hadrons in relativistic heavy-ion collisions are 
originated from a thermal source of a given thermal parameters of temperature 
and baryonic chemical potential \cite{cley}. In \cite{peter} a parametrization 
of the temperature and baryonic chemical potential as a function of $\sqrt
{s_{NN}}$ (in GeV) is obtained from the fits of calculated particle yields and 
ratios to the experimental data at mid-rapidity over a broad energy range of 
$\sqrt{s_{NN}}$=2.7-200 GeV. Then they extend above parametrization to the LHC 
energy and provide a quantitative prediction for LHC $Pb+Pb$ experiments.

Similarly, the transport models are also unable to be in agreement with 
the experimental data of nucleus-nucleus collisions in a broad energy range  
from AGS to SPS, RHIC, and even to LHC without adjusting any parameters. Of 
course, the less number of adjusting parameters the better the model is. In this 
letter the parton and hadron cascade model, PACIAE, is used to simulate the 
0-6\%  (15-25 and 35-45\%) most central $Au+Au$ collisions at $\sqrt{s_{NN}}$=
19.6, 62.4, 130, and 200 GeV and the 0-10\% most central $Pb+Pb$ collisions at 
$\sqrt{s_{NN}}$=5500 GeV. The charged multiplicity and the charged particle 
transverse momentum distribution, pseudorapidity distribution, and extended 
longitudinal scaling are calculated. In the calculations we fix all model 
parameters, except the parameter $b$ in Lund string fragmentation function 
\cite{sjo} 
\begin{equation}
f(z)\propto z^{-1}(1-z)^aexp(-bm_T^2/z).
\label{str}
\end{equation} 
This function expresses the probability that a given $z$ is picked, here $z$ 
refers to the fraction of string energy (momentum) taken away by the produced 
hadron. In Eq. (\ref{str}) the $m_T =\sqrt{p_T^2+m_h^2}$ is transverse mass of 
hadron, $p_T$ and $m_h$ are, respectively, the transverse momentum and rest 
mass of hadron, and the $a$ and $b$ are parameters. As mentioned in \cite{sa2} 
that the parameter $b$ is varied with string density and the higher reaction 
energy (temperature) is corresponding to the lager $b$. According to the 
experimental facts that the temperature of fireball, in relativistic heavy-ion 
collisions, as a function of $\sqrt{s_{NN}}$ increases dramatically first and 
then approaching saturation gradually \cite{brat}, we assume $b$ to be 
approximately a function of $\sqrt{s_{NN}}$ as 
\begin{equation*}
b=b_0\frac{\sqrt{s_{NN}}}{200}, \hspace{0.3cm} if \hspace{0.2cm} \sqrt{s_{NN}}
  \leq 200 \hspace{0.1in} \rm{GeV},
\end{equation*}
\begin{equation}
b=b_0[1+(1-\frac{200}{\sqrt{s_{NN}}})], \hspace{0.3cm} if \hspace{0.2cm} \sqrt
  {s_{NN}}>200 \hspace{0.1in} \rm{GeV},
\end{equation}
where $b_0$=6. As the calculated above observables in $Au+Au$ collisions are in  
good agreement with PHOBOS data (see later), the calculated those observables in 
$Pb+Pb$ collisions are thus a reliable prediction.  
\begin{center}
\begin{table}[htbp]
\scriptsize      
\caption{Charged multiplicity in $Au+Au$ collisions at $\sqrt{s_{NN}}$
         =19.6, 62.4, 130, and 200 GeV and in $Pb+Pb$ collisions at $\sqrt{s_
         {NN}}$=5500 GeV.}
\begin{tabular}{cccccc}
\hline\hline
      & \multicolumn{4}{c}{Au+Au}&  Pb+Pb  \\
\cline{2-5}
Energy & 19.6& 62.4& 130& 200& 5500 \\
(GeV)  & & & & & \\
\cline{2-5}
Centrality& \multicolumn{4}{c}{0-6\%}&  0-10\% \\
Multiplicity & & & & & \\
Exp.   & 1689$\pm$100$^1$& 2845$\pm$142$^2$& 4170$\pm$210$^1$& 5060$\pm$250$^1$
& \\
PACIAE & 1533& 2919& 4140& 5001& 14695 \\
$\frac{dN_{ch}}{d\eta}|_{\eta =0}$ & 250$^3$& 425$^3$& 559$^3$& 641$^3$& 
 1196$^4$ \\
b  & 0.58& 2& 4& 6 & 12 \\
\hline\hline
\multicolumn{6}{l}{$^1$ PHOBOS data taken from \cite{phob1}.}\\
\multicolumn{6}{l}{$^2$ PHOBOS data taken from \cite{phob2}.}\\
\multicolumn{6}{l}{$^3$ Estimated from calculations in 0.2$<\eta <$1.4 .}\\
\multicolumn{6}{l}{$^4$ Estimated from calculations in $|\eta|<$0.12 .}\\
\end{tabular}
\label{mul}
\end{table}
\end{center}
\normalsize
The parton and hadron cascade model, PACIAE \cite{sa}, is based on PYTHIA (the 
model for hadron-hadron collision) \cite{sjo} and is composed of four stages: 
the parton initialization, parton evolution (scattering), hadronization, and 
hadron evolution (rescattering). 

\begin{enumerate}
\item Parton initialization: \\
In the PACIAE model the nucleus-nucleus collision is decomposed into nucleon
-nucleon (NN) collisions according to the geometry of nucleus-nucleus collision. 
A NN collision is performed by the PYTHIA model \cite{sjo} with the 
hadronization process switched-off. Thus the consequence of a nucleus-nucleus 
collision is a state composed of quark pairs, diquark pairs, gluons, and very 
few hadronic remnants. If the diquark (anti-diquark) is split forcedly into 
quarks (anti-quarks) randomly, the partonic initial state of a nucleus-nucleus 
collision is reached. \\ \\

\begin{figure}
\centerline{\hspace{-0.1in}
\epsfig{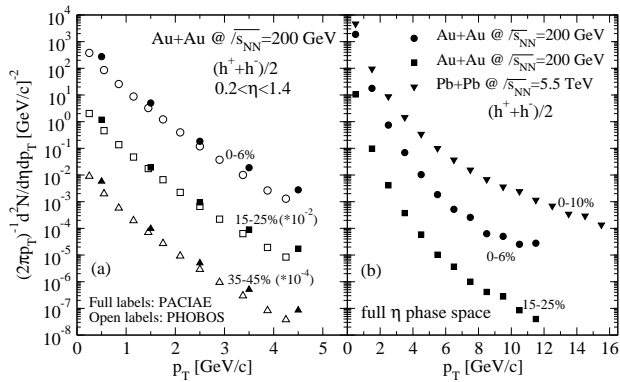}}
\vspace{-0.1in}
\caption{The charged particle transverse momentum distributions in $Au+Au$ 
collisions at $\sqrt{s_{NN}}$=200 GeV and $Pb+Pb$ collisions at $\sqrt
{s_{NN}}$=5500 GeV. PHOBOS data are taken from \cite{phob3}.}
\label{aupb_pt}
\end{figure} 
\item Parton evolution (scattering):\\
The parton initialization stage is followed by parton evolution (scattering). 
Here the $2\rightarrow 2$ LO-pQCD differential cross sections \cite{comb} are
employed. The differential cross section of a sub-process $ij\rightarrow kl$ 
reads
\begin{equation}
\frac{d\sigma_{ij\rightarrow
kl}}{d\hat{t}}=K\frac{\pi\alpha_s^2}{\hat{s}}\sum_{ij\rightarrow kl},
\end{equation}
where the factor $K$ is introduced counting for the higher order pQCD and 
non-perturbative QCD corrections, $\alpha_s$ stands for the strong 
(running) coupling constant, and $\hat{s}$, $\hat{t}$, and $\hat{u}$
are the Mandelstam variables. For the process $q_1q_2 \rightarrow q_1q_2$, 
for instance, one has
\begin{equation}
\sum_{q_1q_2\rightarrow
q_1q_2}=\frac{4}{9}\frac{\hat{s}^2+\hat{u}^2}{\hat{t}^2}.
\label{eq3}
\end{equation}
It diverges at $\hat{t}$=0 and has to be regularized by introducing the 
parton colour screen mass $\mu$ as follows
\begin{equation}
\sum_{q_1q_2\rightarrow
q_1q_2}=\frac{4}{9}\frac{\hat{s}^2+\hat{u}^2}{(\hat{t}-\mu^2)^2}.
\end{equation}
The total cross section of the parton collision $i+j$ is then
\begin{equation}
\sigma_{ij}(\hat{s})=\sum_{k,l}\int_{-\hat{s}}^{0}d\hat{t}
\frac{d\sigma_{ij\to kl}}{d\hat{t}}.
\end{equation}
With above total and differential cross sections the parton evolution (parton
scattering) can be simulated by the Monte Carlo method.
\item Hadronization: \\
The hadronization at the moment of partonic freeze-out (no more partonic  
collision at all) is consequent on the parton evolution stage. In the PACIAE 
model, the phenomenological fragmentation model (String Fragmentation, 
Independent Fragmentation, or Cluster Fragmentation) \cite{sjo} and the 
coalescence model are supplied for hadronization of partons after scattering. 
The String Fragmentation model is adopted in this letter. We refer to \cite{sa} 
for the details of the hadronization stage.
\begin{widetext}
\begin{center}
\begin{figure}
\centerline{\hspace{-0.5in}
\epsfig{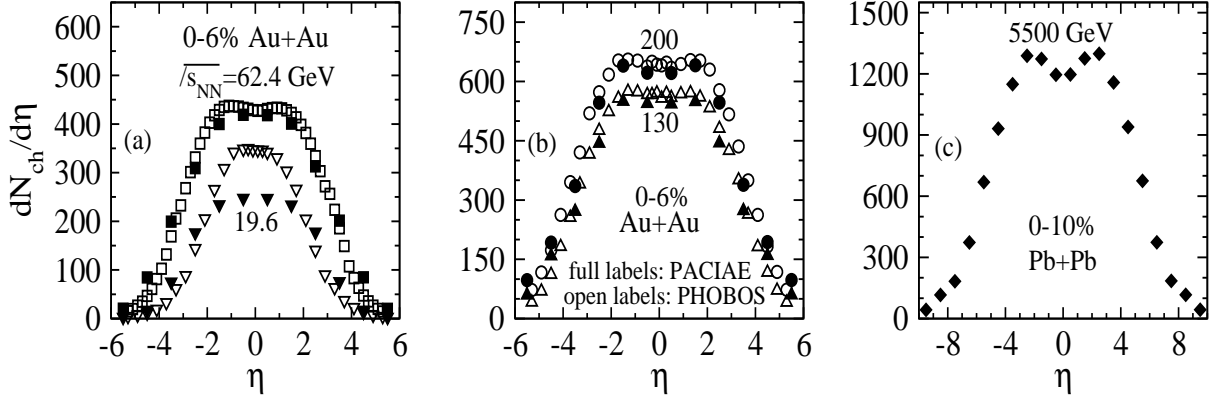}}
\vspace{-0.1in}
\caption{The charged particle pseudorapidity distributions in 0-6\% most 
central $Au+Au$ collisions at $\sqrt{s_{NN}}$=19.6, 62.4, 130, and 200 GeV 
and in 0-10\% most central $Pb+Pb$ collisions at $\sqrt{s_{NN}}$=5500 GeV. 
The full labels are calculated results by the PACIAE model and the open labels 
are the corresponding PHOBOS data. PHOBOS data are taken from \cite{phob1} (for 
$\sqrt{s_{NN}}$=19.6, 130, and 200 GeV) and \cite{phob2} (for $\sqrt{s_{NN}}$
=62.4 GeV).}
\label{aupb_eta}
\end{figure}
\end{center}
\end{widetext}
\item Hadron evolution (rescattering): \\
After hadronization the rescattering among produced hadrons is dealt with the
usual two-body collision model. The details of hadronic rescattering can see 
\cite{sa1}.
\end{enumerate}

We compare the calculated charged multiplicity with the PHOBOS data (taken from 
\cite{phob1,phob2}) in 0-6\% most central $Au+Au$ collisions at $\sqrt{s_{NN}}$
=19.6, 62.4, 130, and 200 GeV in Table \ref{mul}. One sees here that the  
PHOBOS data are well reproduced within error bars by the PACIAE  
calculations, except the case of $\sqrt{s_{NN}}$=19.6 GeV. That is consistent 
with the fact that the PYTHIA model is more suitable for higher reaction energy. 
The predicted charged multiplicity and charged particle pseudorapidity density 
at mid-pseudorapidity in 0-10\% most central $Pb+Pb$ collisions at 
$\sqrt{s_{NN}}$=5500 GeV is also given in this table. That pseudorapidity 
density, $\frac{dN_{ch}}{d\eta}|_{\eta =0}\sim $1200, is within the predicted  
values of other sixteen models listed in \cite{arm2}.  

In Fig. \ref{aupb_pt} (a) we compare the calculated charged particle 
transverse momentum distributions with the corresponding PHOBOS data (taken 
from \cite{phob3}) in $Au+Au$ collisions at $\sqrt{s_{NN}}$=200 GeV. Here the 
open circles, squares, and triangles are, respectively, the PHOBOS data of the 
0-6, 15-25, and 35-45\% most central $Au+Au$ collisions and the full labels are 
the PACIAE model results. We see in panel (a) that the PHOBOS data are 
reasonably good reproduced. The prediction for charged particle transverse 
momentum distribution (triangles-down) in 0-10\% most central $Pb+Pb$ collisions 
at $\sqrt{s_{NN}}$=5500 GeV are given in Fig. \ref{aupb_pt} (b). For comparison,  
the charged particle transverse momentum distributions in 0-6 (circles) and 
15-25\% (squares) most central $Au+Au$ at $\sqrt{s_{NN}}$=200 GeV  are also given 
in panel (b). Note that the charged particle transverse momentum distributions 
shown in Fig. \ref{aupb_pt} (b) is integrated over full pseudorapidity. 

Figure \ref{aupb_eta} shows the charged particle pseudorapidity distributions 
in 0-6\% most central $Au+Au$ collisions at $\sqrt{s_{NN}}$=19.6 and 62.4 GeV 
(panel (a)) and at $\sqrt{s_{NN}}$=130 and 200 GeV (panel (b)) and in 0-10\% 
most central $Pb+Pb$ collisions at $\sqrt{s_{NN}}$=5500 GeV (panel (c)). In this 
figure the open labels are the PHOBOS data and the full labels are the PACIAE  
model results. The squares and triangles-down in panel (a) are, respectively, for 
$\sqrt{s_{NN}}$=62.4 and 19.6 GeV and the circles and triangles-up in panel (b) 
are, respectively, for $\sqrt{s_{NN}}$=200 and 130 GeV. One sees again in panels 
(a) and (b) that the PHOBOS data are reasonably good reproduced, except the case 
of $\sqrt{s_{NN}}$=19.6 GeV. Fig. \ref{aupb_eta} (c) gives the PACIAE model 
prediction for the charged particle pseudorapidity distribution in 0-10\% 
most central $Pb+Pb$ collisions at $\sqrt{s_{NN}}$=5500 GeV. In panel (c) one 
sees that there is a deep valley at the mid-pseudorapidity.    

\begin{figure}[h]
\centerline{\hspace{-0.1in}
\epsfig{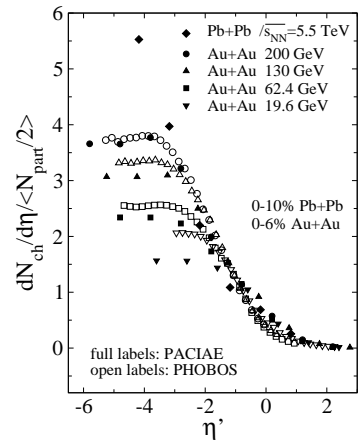}}
%\vspace{0.2in}
\vspace{-0.1in}
\caption{The shifted charged particle pseudorapidity distributions in 0-6\% 
most central $Au+Au$ collisions at $\sqrt{s_{NN}}$=19.6, 62.4, 130, and 200 
GeV and in 0-10\% most central $Pb+Pb$ collisions at $\sqrt{s_{NN}}$=5500 
GeV. The shifted pseudorapidity $\eta^{\prime}=\eta -y_{beam}$, where $y_{beam}$ 
is the beam rapidity. And the $N_{part}$ refers to the number of participant 
nucleons.}
\label{aupb_sca}
\end{figure}

In Fig. \ref{aupb_sca} we give the shifted charged particle pseudorapidity 
distributions in 0-6\% most central $Au+Au$ collisions at $\sqrt{s_{NN}}$=
19.6 (triangles-down), 62.4 (squares), 130 (triangles-up), and 200 GeV 
(circles) and in 0-10\% most central $Pb+Pb$ collisions at $\sqrt{s_{NN}}$=5500 
GeV (diamonds). Here the shifted pseudorapidity is
\begin{equation}
\eta^{\prime}=\eta-y_{beam},
\end{equation} 
where $y_{beam}$ is the beam rapidity. We see in  Fig. \ref{aupb_sca} that 
the extended longitudinal scaling \cite{phob2} is well kept not only among 
$Au+Au$ collisions at a variety values of $\sqrt{s_{NN}}$ but also among 
$Au+Au$ and $Pb+Pb$ collisions. \\

In summary, we have used the parton and hadron cascade model PACIAE to 
simulate the 0-6, 15-25, and 35-45\% most central $Au+Au$ collisions at 
$\sqrt{s_{NN}}$=19.6, 62.4, 130, and 200 GeV and the 0-10\% most central 
$Pb+Pb$ collisions at $\sqrt{s_{NN}}$=5500 GeV. The charged multiplicity 
and the charged particle transverse momentum distribution, pseudorapidity 
distribution, and extended longitudinal scaling are calculated. For $Au+Au$ 
collisions the calculated above observables are in good agreement with the 
corresponding PHOBOS data. Thus the above observables calculated for $Pb+Pb$ 
collisions would be a reliable prediction.\\  

The financial support from NSFC (10475032, 10605040, and 10635020)
in China are acknowledged.

\end{document}